# Simulation of Muon Background at the ILC*

L. Keller and G. White

*SLAC National Accelerator Laboratory, 2575 Sand Hill Road, Menlo Park, CA 94025*

## Abstract

Beginning with the first linear collider, SLC at SLAC, it was quickly discovered that high energy muons that are produced in halo collimators in the beam delivery system can cause a significant background in the experiment detector. Following publication of the ILC Technical Design Report, May 2013 [1] a simulation of this background has been made using simulation codes MUCARLO and GEANT4. It became clear that to mitigate this background, various magnetic devices were going to be needed.

## Introduction

When the Mark II detector first turned on at the SLC in 1987, it was flooded with an unanticipated, intolerable muon background. It was quickly discovered that the muons were coming from a tungsten halo collimator at the beginning of the final focus (beam delivery) system about 170m from the IP. If the halo collimator was opened, then synchrotron radiation from halo in the final doublet began hitting the IP beampipe, also causing an intolerable background in the vertex detector. A FORTRAN code, MUCARLO [2], was quickly written by G. Feldman in order to simulate muon background in the Mark II and study where to locate magnetic spoilers to reduce it to a tolerable level. The program successfully reproduced the experimental results, both before and after the magnetic spoilers were installed. Over the years MUCARLO has been expanded and has been used at SLAC in muon shielding designs for SLD, radiation protection, fixed target experiments, and in muon background estimates for the NLC and ILC. This work shows the results of the background simulations using the parameters in the ILC Technical Design Report.

*Work supported under US Department of Energy contract DE-AC02-76SF00515
*Talk presented at the International Workshop on Future Linear Colliders (LCWS2018), Arlington, Texas, 22-26 October 2018. C18-10-22*

# Muon Production

The muon production calculation in MUCARLO has been described previously in SLAC-Pub-5533 [3] and SLAC-Pub-6385 [4] and will be summarized here. There are two production mechanisms, Bethe-Heitler and direct annihilation. Bethe-Heitler is a two-stage process where the beam electron or positron radiates a bremsstrahlung photon which produces a muon pair off a target nucleus. For thick targets like absorbers and protection collimators the photon track length vs. photon energy is estimated by the Clement-Kessler shower approximation. For thin targets like spoilers the photon track length is simulated in Program FLUKA [5]. The differential cross section for photon-nucleus muon production is given by Y. Tsai [6]. Direct annihilation occurs when positrons interact with atomic electrons to produce a muon pair. The differential cross section for positron-electron annihilation is given by Brodsky [7]. The laboratory energy threshold for this process is a positron energy of 43.7 GeV. The shower track lengths vs. energy from the positron beam in the spoilers and thick targets is also simulated with FLUKA. Figure 1 shows the muon yield from a 250 GeV positron beam impacting a thick HiZ copper target, corresponding to protection collimators, and a thin copper target, corresponding to halo spoilers. It is seen that the muon yield is 1-2 orders of magnitude lower in both the Bethe-Heitler and the direct annihilation process except near the highest muon energies where direct annihilation in spoilers is more probable that Bethe-Heitler in spoilers. This is important because, as will be seen later, the highest energy muons have a much better chance of reaching the detector.

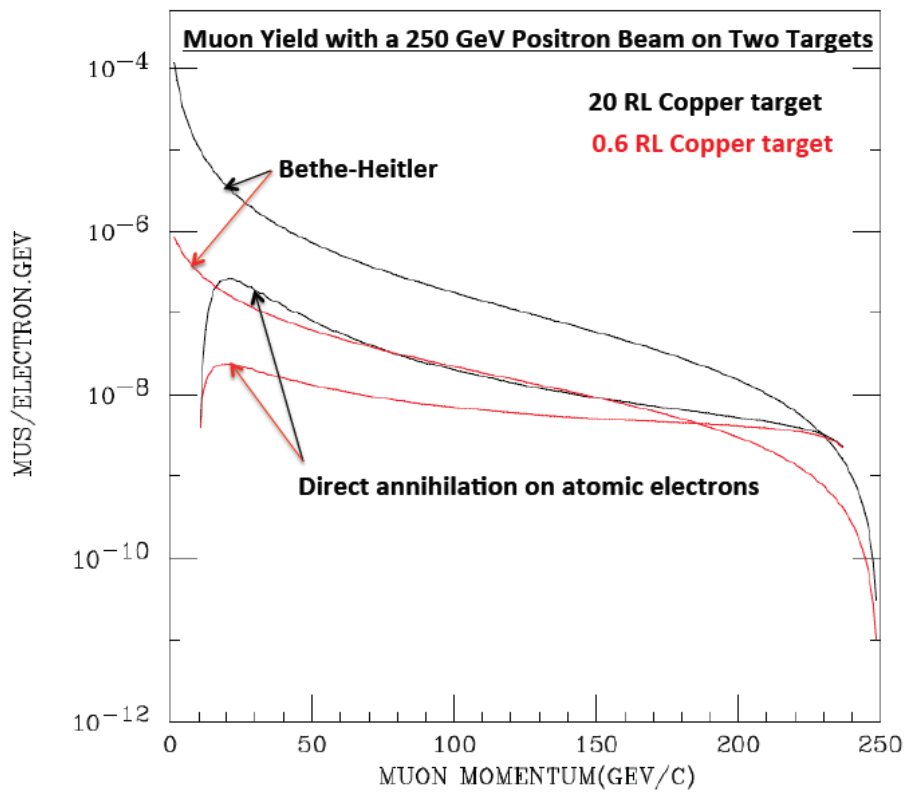

*Figure 1. Muon yield from a 250 GeV positron beam impacting two different copper targets.*

# Beam Delivery System and Muon Sources

The beam delivery system begins at the exit of the linac and transports the electron and positron beams to the interaction point (IP) with a crossing angle of 14 mrad (7 mrad each side). The crossing angle allows the spent beams to enter extraction lines leading to the high-power beam dumps. Beginning 2254 m from the IP, the system consists of the following sections leading to the IP: machine protection collimators, skew correction and emittance diagnostics, polarimeter, fast kickers for the tuneup dump beamline, betatron collimation, energy collimation, final transformer, and final doublet. In the betatron section there are two spoiler-absorber combinations; these are the main muon sources. The copper spoilers are 0.6 rl thick with adjustable gaps in both planes. Their purpose is to intercept beam halo, which would otherwise cause synchrotron radiation from the final doublet to impact the IP beampipe. The spoilers must be relatively thin to survive the power deposition from the halo as well as survive a direct hit from a slightly mis-steered single bunch. The thin spoilers create an electromagnetic shower, which is absorbed in thick protection collimators and dedicated absorbers down-beam from each spoiler. Figure 2 is a plan view of the tunnel in the MUCARLO and GEANT4 [8] models. It shows the muon sources in black and the five magnetized toroids and tunnel-filler muon spoiler in red.

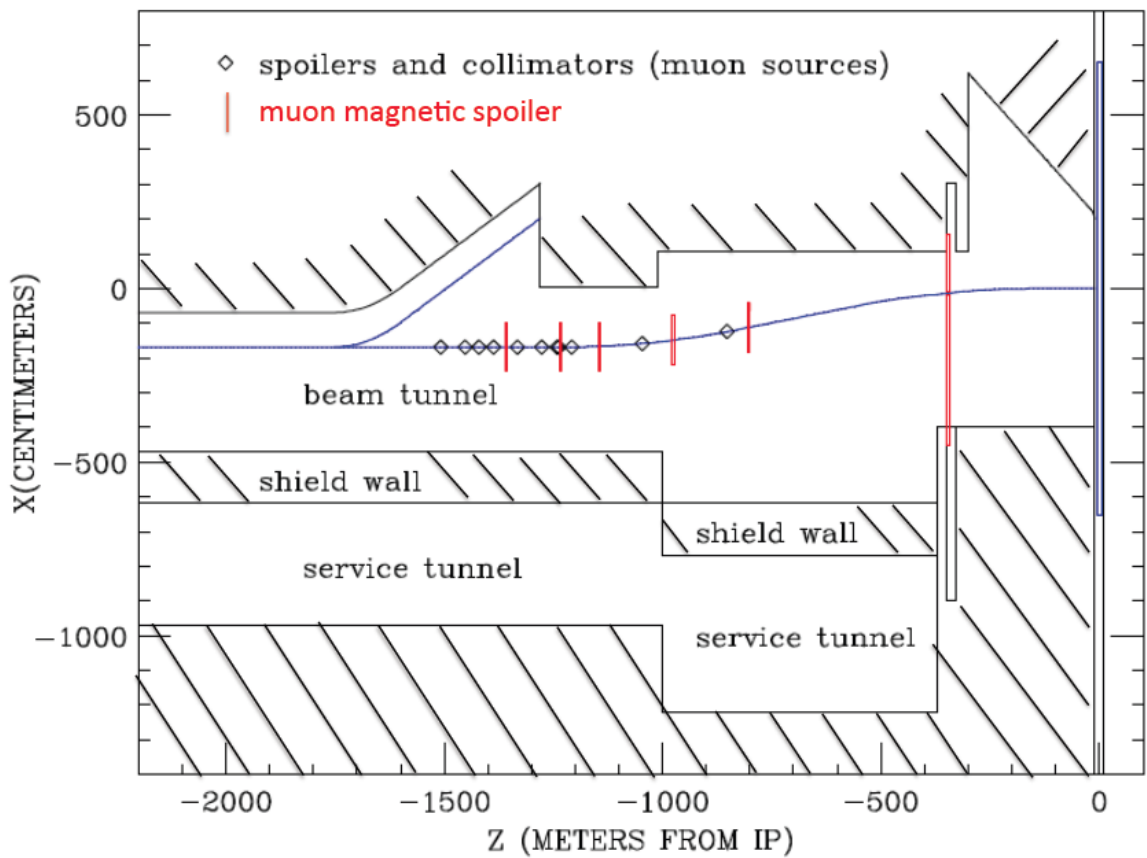

*Figure 2. Plan view of the tunnel in the MUCARLO model. The beam axis is shown in blue. The first muon source, SP2, is located 1508m from the IP. It is inboard from the branch line to the tuneup dump which begins at about 1700m. The red lines show the location of cylindrical, magnetized spoilers and the magnetized steel wall at 350m. The vertical blue rectangle at the IP represents a 6.5m radius detector. The 7mrad crossing angle is not included in this picture.*

The two betatron section beam halo spoilers, SP2 and SP4, are located at 1508 m and 1322 m from the IP, and their gaps are each set to 6.4 $\sigma_x$ (930µm) and 44 $\sigma_y$ (400µm). These two spoilers intercept 100% of the beam halo. There is a 3rd spoiler, SPEX, located at the high dispersion point in the energy collimation section, whose gap is set to 1% $\Delta P/P$. Since the energy spread in the main beam is only about 0.2%, this spoiler receives no halo. It mainly acts as a protection spoiler in case a klystron failure lets off-energy bunches into the beam delivery system. All other muon sources are thick, HiZ protection collimators and absorbers which protect the 1.0 cm radius betatron and energy section quadrupoles from the spoiler showers. The copper protection collimators have 0.5cm radius and the copper absorbers are adjustable 4-jaw collimators with ±0.4cm gaps in both planes. Figure 3 shows two views of a magnetized toroid spoiler,

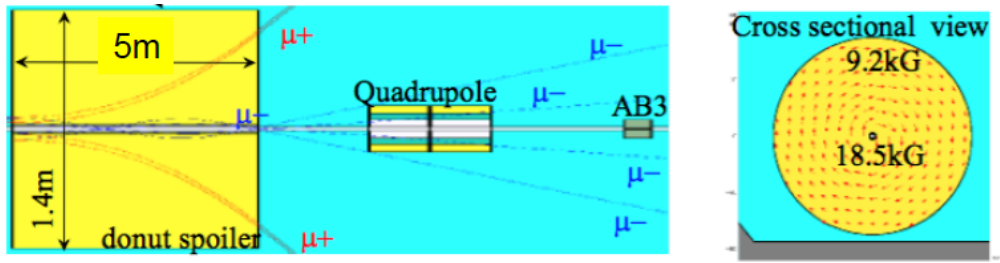

*Figure 3. The 5m long, 70cm radius magnetized toroid spoilers. The beam hole has a 1 cm radius and a magnetic shield to keep the beam hole free of magnetic field. They weigh about 60 metric tons each. As seen in the picture the toroids defocus one sign muon charge and focus the other sign, which can channel the muon along the beam axis. Therefore the spoiler location needs to be near quadrupoles or dipoles which can also deflect the muon into the tunnel walls.*

The beamline elements are close to one wall leaving room for an aisle on the other side of the beamline. The model also contains a service tunnel which is separated by a 1.5m thick concrete wall parallel to the beam tunnel. The figure shows the electron tunnel, which also has room on one side for the low energy positron beam coming from the positron source at the end of the linac. The positron beam delivery tunnel (not shown) has room for the electron source linac in a wide region starting about 1000m from the IP. In the region 355-340m from the IP is a wide cavity in which a magnetized steel wall can be located in case that is necessary to reduce the number of muons reaching the IP. This model has a 5m long magnetized wall for some of the runs. Starting 300m from the IP is the extraction line area for the spent beam coming from the opposite direction. In the model the beam elements are located 1.1m from the tunnel floor, and the tunnel vertical height is 5m along the entire beam line. In practice the tunnel would have rounded corners near the ceiling, but these are not included the model. Figure 4 shows two views of the magnetized wall which fills the tunnel at 350m from the IP.

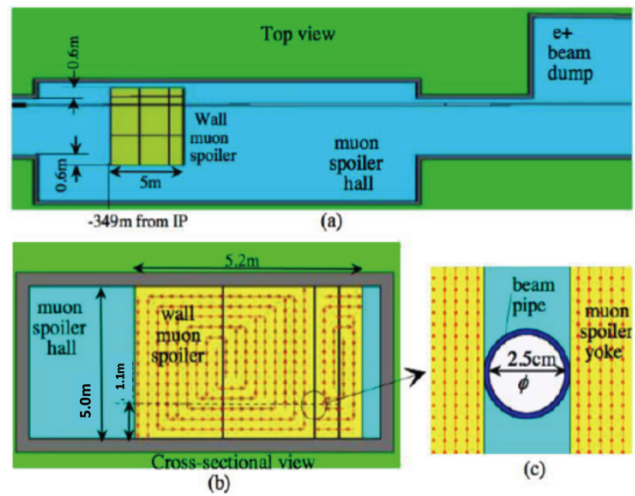

*Figure 4. (a) Plan view location of the 5m long magnetized wall which consists of two pieces on either side of the beam axis (b) Beam view showing how the spoiler blocks the 5m wide tunnel but leaves room for passage around it. (c) The beampipe diameter is 2.5cm and is surrounded by a bucking coil and soft iron shielding to keep the beam region field free. The tunnel cavity allows the wall to be lengthened along the beam in case 5m is not enough deflection.*

## Conditions and Steps in the Simulation

Listed below are the conditions and steps for the calculation.

1) Spoilers SP2 and SP4 are set to 930μ (6.4 $\sigma_x$) and 400μ (44 $\sigma_y$). Energy spoiler SPEX is set to 1% δP/P. Spoilers are 0.6 rl copper.
2) Absorbers are 4-jaw rectangular ±0.4cm gaps, 20 rl copper.
3) Protection collimators are round, 0.5cm radius gap, 20 rl copper.
4) Use halo program, DISTR.f [9] to generate 1/R trajectories with 5-13 $\sigma_x$ and 36-93 $\sigma_y$ and δP/P = 0.19%. (Figure 5 below). Intercepted halo is defined to be 0.1% of the bunch intensity, i.e. $2 \times 10^7$/bunch.
5) Input $1 \times 10^5$ halo trajectories to TURTLE [10] which records the fraction of hits on the spoilers (SP's), absorbers (AB's) and protection collimators (PC's) as well as the maximum energy of the electromagnetic shower impacting the absorbers and protection collimators.
6) Use MUCARLO to produce and track muons from 0.6 rl SP's and 20 rl AB's and PC's.
7) As a check use Lucretia [11] halo particle tracking with built-in GEANT4 model interface.
8) Need four separate MUCARLO decks because the production rates differ for positron and electron beams and for 0.6 rl and 20 rl copper targets.
9) Use the same tunnel model for both positron and electron beams.
10) The tunnel ends 10m from the IP. Muons which reach 10m are transported to the IP through air, i.e. there are no detector components in the model.
11) Record the positive and negative muon hits/bunch in a 6.5m radius detector for three muon spoiler conditions:

    a) No muon spoilers.

b) Five, 5m long, 70cm outside radius toroids located 1356m, 1231m, 1143m, 973m, and 800m from the IP. Two of the toroids are located within the betatron collimation section, two in the energy collimation section, and one toroid is located just inboard from the energy collimation section.
c) In addition to the five toroid spoilers, 5m long rectangular spoilers which completely fill the tunnel and extend beyond the left and right walls by 0.5m, and are located 344m from the IP.

For each muon reaching 10m from the IP and each spoiler condition, record its 4-vector and time difference with respect to the bunch crossing. These trajectories can then be used by the detector designers to see the effect on the various detector elements.

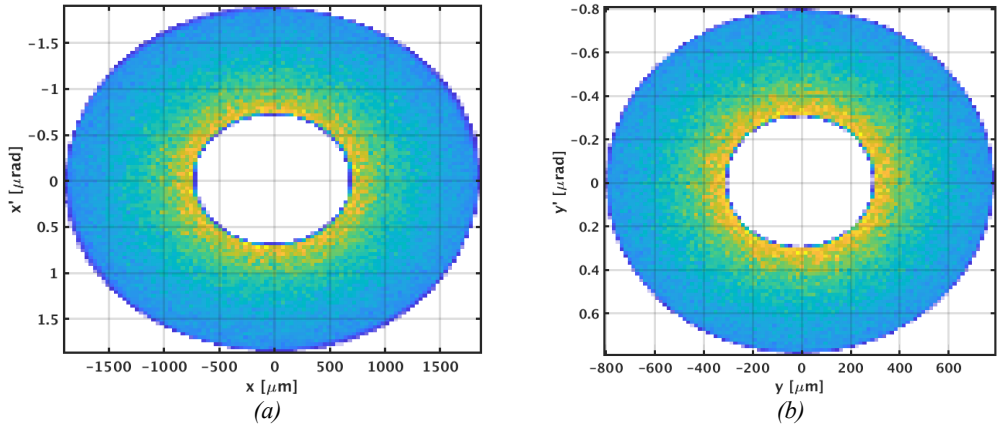

Figure 5. (a) X/X' phase space or 5-13 $\sigma_x$ generated beam halo. The horizontal gap is 6.4 $\sigma_x$ (930µ) half-gap. (b) Y/Y' phase space for 36-93 $\sigma_y$ generated beam halo. The vertical half-gap is 44 $\sigma_y$ (400µ).

Table 1 shows the fractional distribution of 250 GeV halo trajectories hitting spoilers SP2 and SP4, and the resulting hit fraction on absorbers and protection collimators, as well as the maximum energy of electromagnetic shower particles impacting the HiZ collimators and absorbers. The primary beam hits only spoilers resulting in secondary hits on absorbers and protection collimators.

Table 1: Fractional Distribution of Hits on Spoilers, Absorbers, and Protection Collimators.

| Element | Distance from IP (m) | Hit Fraction | Max. Beam Energy (GeV) |
|---|---|---|---|
| SP2 | 1508 | 0.654 | 250 |
| PC1 | 1452 | 0.237 | 35 |
| AB3 | 1420 | 0.192 | 150 |
| PC2 | 1387 | 0.171 | 150 |
| SP4 | 1332 | 0.337 | 250 |
| PC5 | 1276 | 0.117 | 75 |
| PC5A | 1242 | 0.083 | 200 |
| AB5 | 1237 | 0.030 | 150 |
| PC6 | 1208 | 0.022 | 200 |
| PC7 | 1047 | 0.042 | 250 |
| ABE | 852 | 0.011 | 250 |

The halo spoilers SP2 and SP4 together intercept 100% of the halo, while the total hit fraction on PC's and AB's is 90.6%. Another 3.7% of lost particles are spread along the entire beam line in quadrupoles and dipoles and are not included by MUCARLO in muon production. The remaining 5.7% of lost particles are considered by TURTLE to be lost in spoilers SP2 and SP4. The MUCARLO simulation uses the maximum energy of the electrons and positrons hitting the HiZ collimators and will therefore give an overestimate of muons produced.

## Results

Table 2 shows the primary result of the MUCARLO and GEANT4 simulations at two center of mass ILC energies.

*Table 2: Number/bunch crossing in a 6.5m radius detector.*

| Tunnel Condition | ILC500 | | | | ILC250 | |
| --- | --- | --- | --- | --- | --- | --- |
| | Detector | | TPC* | | Detector | TPC* |
| | MUCARLO | GEANT4 | MUCARLO | GEANT4 | MUCARLO | MUCARLO |
| **No Spoilers** | 130 | 39 | 7920 | 4000 | 38 | 2740 |
| **Five, 5m toroid spoilers** | 4.3 | 2.8 | 397 | 280 | 1.3 | 119 |
| **Five, 5m toroid spoilers & one 5m spoiler (z=344-349m) fills tunnel** | 0.6 | -- | 23 | -- | 0.03 | 2 |

*TPC: 2.5m radius, 200 bunches, sensitive time 110μsec.

It is seen that the GEANT4 results at ILC500 are approximately a factor of 2-3 below MUCARLO. Over the years the MUCARLO simulation has been verified many times, see for example [12] where MARS [13] and MUCARLO simulations were compared in an older version of the ILC tunnel. The GEANT4 results were done as a quick check of MUCARLO: there is limited statistics available: only ~150 simulated muons in the case with spoilers are produced at the detector compared with 20M simulated muons from MUCARLO. The final results are thus subject to considerable statistical error. The GEANT4 model is more general and is used to understand the overall background conditions in the tunnel and includes multiple physics processes, hence the final muon statistics were limited by the available compute time for this study as the muon production rates are heavily suppressed compared to the other processes being studied. A further refinement to improve the muon statistics in the future will require the application of weighting techniques to enhance the muon production and will be the subject of a future study. In the table the TPC is defined as having a 2.5m radius and a sensitive time of 110μsec in 200 bunches. Many muons are out-of-time with respect to the bunch crossing depending on the time to reach the detector and which detector elements are hit.

Figures 6 to 12 show details of muons which reach the detector for different spoiler and beam energy conditions. Figures 6 and 7 are examples of muon trajectories in the horizontal and vertical plane which reach the detector from spoilers SP2 and SP4 on the electron beam side. In this case there were only 3 magnetized toroid spoilers and the 5m magnetized wall. The

example shows how nearby beam line magnets influence the charge and energy of muons that reach the IP

Green = positive, red = negative.

The toriod spoilers defocus negative charge, so almost all muons reaching the detector are positively charged.

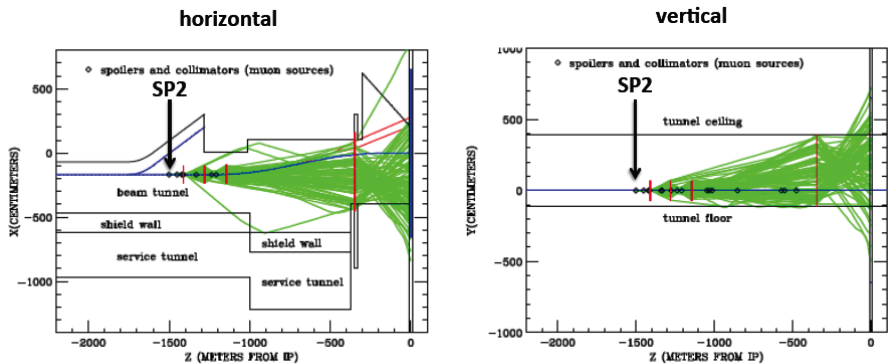

Figure 6. Examples of muon trajectories which reach the IP from halo on SP2 on the electron beam side. In this case the toroid spoilers focus positive charge so very few negative muons reach the detector.

Green = positive, red = negative.

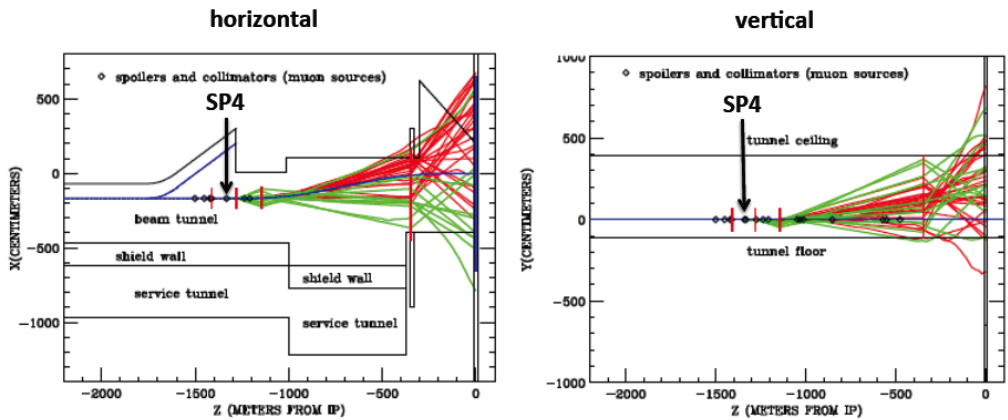

Figure 7. Examples of muon trajectories which reach the IP from halo on SP4 on the electron beam side. Since SP4 and the toroid spoilers are closer to the energy collimation section, even negative sign muons can reach the IP.

Figure 8 shows how muons are concentrated in the 6.5m radius for the condition of 5 toroid spoilers and the 5m magnetized wall. Only those muons which hit the detector are shown in the plot.

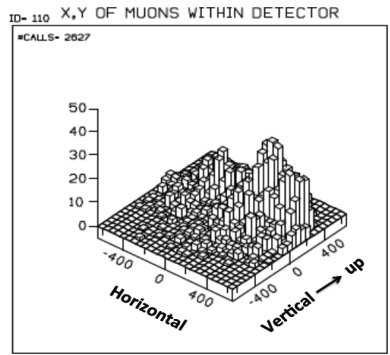
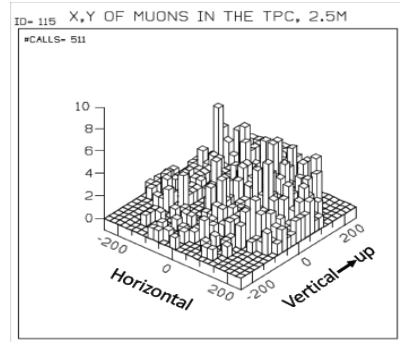

*(a)* *(b)*

*Figure 8. (a) Distribution of muons which reach the 6.5m radius detector for ILC500 with 5 toroid spoilers and the 5m magnetized wall. (b) Distribution of muons which impact a 2.5m radius TPC for ILC500 with 5 toroid spoilers and the 5m magnetized wall. There is naturally a concentration above the tunnel floor and on the aisle side of the tunnel.*

For the conditions of (1) no magnetic spoilers and (2) five magnetized spoilers there is a concentration of muons in the detector above the beam axis and on the aisle side (See Figure 9 below). When the tunnel-filling magnetized wall is added, the muons are much more spread out in and beyond the detector. Figure 9 shows how the spatial distribution of muons at the detector varies with two different spoiler combinations and beam energy.

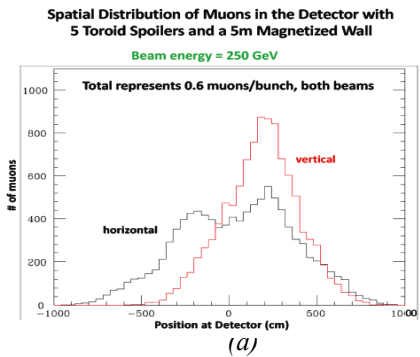
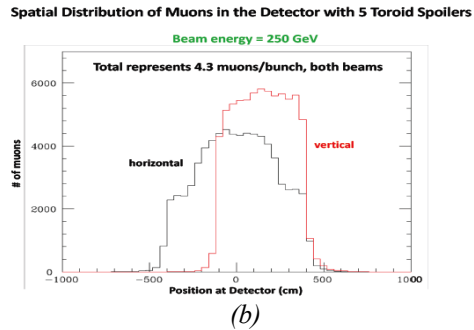

*(a)* *(b)*

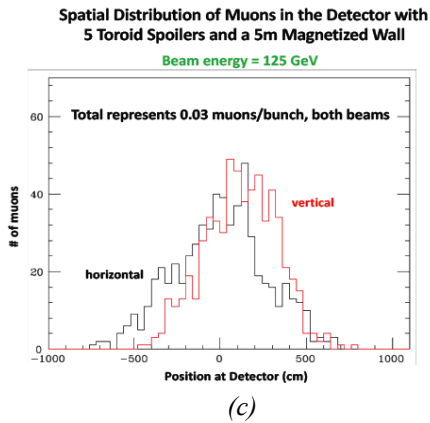
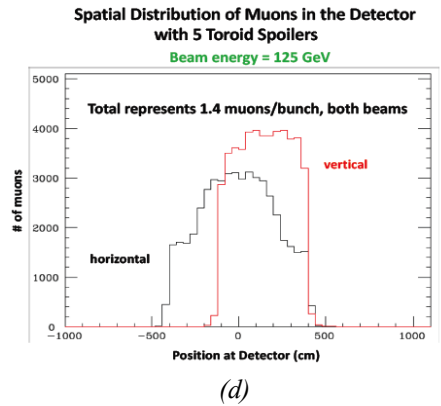

*(c)* *(d)*

*Figure 9. Spatial distribution at detector for ILC500 and ILC250 and for two magnetized spoiler conditions. Note the difference in vertical scale for the same beam halo hitting the betatron spoilers. (a) The magnetized wall has caused some of these muons to miss a 6.5m radius. These are not counted in the 0.6 muons/bunch hitting the detector. (b) For just 5 toroid spoilers the muons are concentrated within the tunnel walls, and all of the muons hit the detector. (c) For ILC250 the spatial distribution is similar to ILC500 except only 0.03muons/bunch hit the detector (d) For ILC250 and 5 toroid spoilers 1,4 muons/bunch hit the detector.*

Figures 10 and 11 show the source momentum of muons that have reached the detector for ILC500 and ILC250. It is seen that for both 250 and 125 GeV beams and 5 toroid spoilers and 5m magnetized wall, that it mostly takes the higher part of the muon spectrum (purple and red) to hit the 6.5m radius detector because the magnetized wall deflects lower momentum muons beyond 6.5m radius. With just the 5 toroid spoilers, (yellow and blue), the situation is reversed, i.e. predominately the lower part of the source spectrum stays within a 6.5m radius.

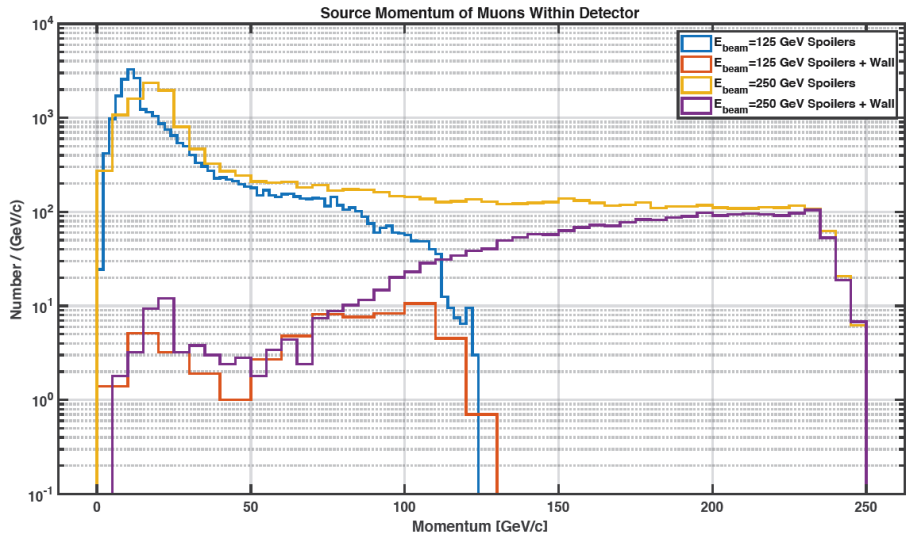

*Figure 10. Source momentum of muons that reach the detector for the condition of 5 toroid spoilers and the 5m magnetized wall for ILC500 and ILC250. Only the higher momentum muons can survive with all the magnetized spoilers (purple curve).*

In Figure 11 with 5 toroid spoilers and 5m wall (purple and red) the final momenta are skewed to higher momentum because the wall has deflected lower momenta more than 6.5m from the beam axis, whereas for only toroid spoilers (yellow and blue), the final momenta are predominantly lower and within the tunnel cross section.

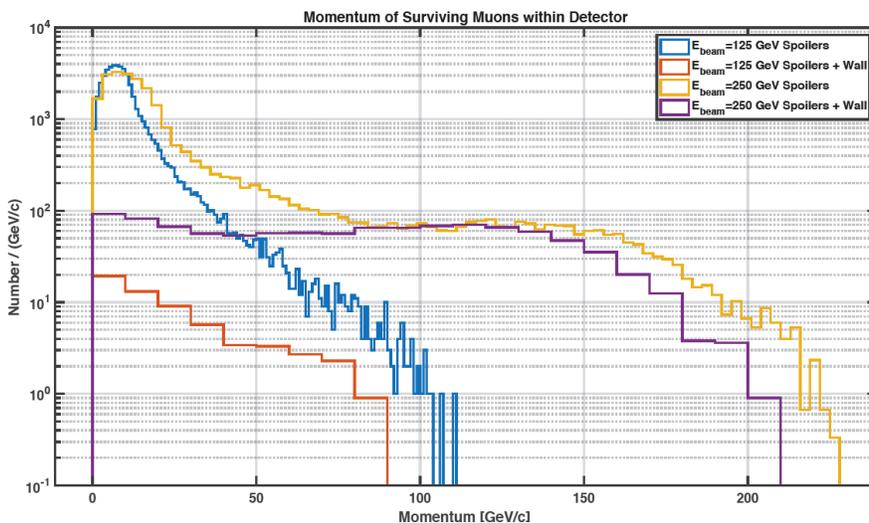

*Figure 11. Final momentum of the same muons as in Figure 9 that have reached the detector for ILC500 and ILC250.*

Figure 12 is an illustration of the muon attenuation factor from all the sources for a 250 GeV positron beam and three spoiler conditions and no spoilers (large attenuation is better). Note that the attenuation from SP2 (1508m) and SP4 (1332m) is at least an order of magnitude smaller for all conditions. This is because with a positron beam more high energy muons are produced in the 0.6 rl spoilers than in the HiZ collimators and absorbers (see Figure 1). Note also that as the source gets closer to the IP, attenuation becomes less. The plot does not take into account the fraction of beam hitting the source (see Table 1).

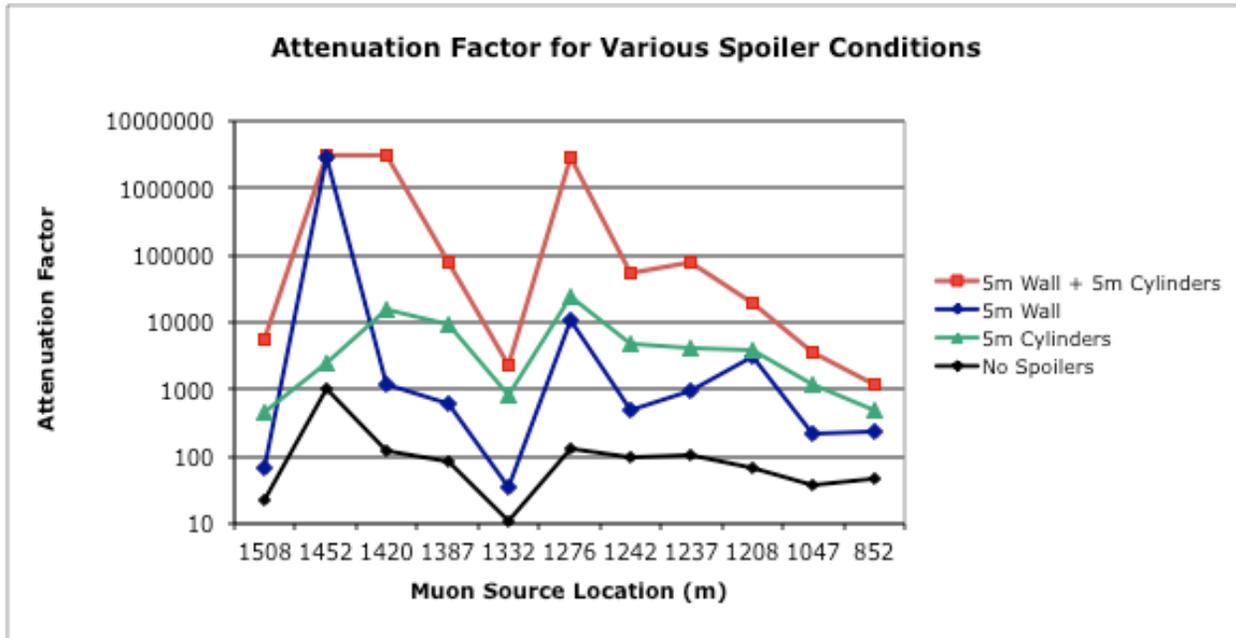

Figure 12. *Illustration of the muon attenuation factor for a 250 GeV positron beam and for three spoiler conditions and no spoilers from all the sources. Large numbers are better.*

In MUCARLO for ILC500 and ILC250, starting with 20M beam halo hits on the betatron spoilers in each beam and two spoiler conditions, 4-vectors of each muon are generated from each source. The spoiler conditions are (1) 5 toroid spoilers and (2) 5 toroid spoilers and the 5m magnetized wall. The 4-vectors of muons reaching 10m from the detector have been given to the two experiment collaborations, SiD (14) and ILD (15), for processing in the detectors. Their conclusions are summarized as part of the following Section.

## Summary and Conclusions

1. For tunnel conditions of no magnetic spoilers and 5 toroid spoilers there is a concentration of muons in the detector above the beam axis and on the aisle side. When the tunnel-filling magnetized wall is added, the muons are much more spread out in the detector.

2. Many muons are out-of-time with respect to the bunch crossing depending on which detector elements are hit.

3. Referring to Figure 12, just the 5m magnetized wall is more effective than 3 magnetized toroids, although it would be much more expensive

4. The conclusion for SiD is that although the occupancy in the vertex and tracker detectors is far below the critical value, the occupancy in the calorimeter endcaps for example almost reaches this critical limit for the shielding without the magnetized wall. It was found that the reason for this is not only the higher muon rate but the muons are concentrated in a small area. Overall the magnetized wall does not seem to be necessary in order to limit the muon occupancy in SiD for both studied center-of-mass energies. However, the wall serves as a tertiary containment device against muons and other machine background particles".

5. The conclusions for ILD are as follows

    (a) Calorimeters: high granularity allows easy identification of muons, minimum ionizing hit energies, many hits are out of time with the bunch crossing by several nsec, Reconstruction not a big problem but may have impact on DAQ system design.
    (b) Silicon trackers: most sensors are parallel to the muons and others (FTD) have a small area.
    (c) TPC: almost all muons are parallel to the drift field so each muon effects only a few readout pads.

# References

[1] Behnke, T. et al. [ILC Collaboration]. International Linear Collider Technical Design Report - Volume 4: Detectors. 2013. arXiv:1306.6329 [physics.accph

[2] S. Rokni, et. al., SLAC-PUB-7054, Nov. 1999.

[3] L. Keller, SLAC-PUB-5533, April 1991.

[4] L. Keller, SLAC-PUB-6385, October 1993.

[5] A. Fasso, A. Ferrari, J. Ranft, and P. R. Sala, "FLUKA": A Multi-particle Transport Code", CERN-2005-10(20050, INFN/TC_05/11, SLC-R-773.

[6] Y. Tsai, Rev. Mod. Phys., Vol 46 No. 4, (1974), p.815

[7] S. J. Brodsky and R. F. Lebed, Phys. Rev.Lett.**102**, 213401, May 2009.

[8] Program GEANT4, Nuclear Instruments and Methods in Physics Research A 506 (2003) 250-303.

[9] I.S. Baishev, et. al., STRUCT Program User's Reference Manual, http://www-ap.fnal.gov/users/drozhdin/

[10] D.C. Carey, et. al., SLAC-246 (1980).

[11] http://www.slac.stanford.edu/accel/ilc/codes/Lucretia/

[12] SLAC-PUB 12741. A.I. Drozhdin, N.V. Mokhov, N. Nakao, S.I. Striganov, Fermilab, Bavia, IL 60510, USA   L. Keller, SLAC, Stanford, CA 94025, USA. Suppression of Muon Background Generated in the ILC Beam Delivery System, August 2007

[13] N.V. Mokhov, "The MARS code System User's Guide", Fermilab-FN-628 (1995); N.V. Mokhov, et. al., Fermilab-Conf. 04/053 (2004); http://www-ap.fnal.gov/MARS/.

[14] A. Schuetz, arXiv:1703.05738v1 [physics.ins-det], 16 Mar 2017.

[15] D. Jeans, KEK, ILD meeting @ ALCW 18, Fukuoka, Japan, June 18, 2018.